\documentclass[12pt]{article}
\usepackage{amsmath}
\usepackage{amsthm}
\usepackage{amsfonts}
\usepackage{color}
\usepackage{graphicx}
\usepackage{float}
\usepackage{subfig}
\usepackage{amsmath}
\usepackage{amsfonts}
\usepackage{multirow}
\usepackage{setspace}
\usepackage{array}
\usepackage{amssymb}
\usepackage{lscape}
\usepackage[authoryear]{natbib}
\usepackage{color}
\usepackage[usenames,dvipsnames]{xcolor}
\usepackage[colorlinks]{hyperref}
\usepackage{rotating}
\usepackage{tikz}
\usepackage{framed}
\usepackage{cleveref}
\usepackage{autonum}
\usepackage{bbm}
\usepackage{subfig} %allows two side by side figs

\usepackage{microtype}%this package improves kerning and use of whitespace. might use [stretch=10,shrink=10] settings if lines are blurred

\usetikzlibrary{backgrounds}
\usetikzlibrary{arrows,shapes}
\usetikzlibrary{tikzmark}
\usetikzlibrary{calc}

\usepackage{tcolorbox}

\usepackage[margin=1in]{geometry}%sets margins (latex default is 1.5 inch for 12 pt docs)
\usepackage{caption}%allows to set margins for captions

\newtheorem{theorem}{Theorem}
\newtheorem{definition}{Definition}
\newtheorem{re-definition}{Re-Definition}

\newtheorem{axiom}{Axiom}

\newtheorem{remark}{Remark}
\newtheorem{example}{Example}

\hypersetup{
    unicode=false,          % non-Latin characters in Acrobat?s bookmarks
    pdftoolbar=true,        % show Acrobat?s toolbar?
    pdfmenubar=true,        % show Acrobat?s menu?
    pdffitwindow=false,     % window fit to page when opened
    pdfstartview={FitH},    % fits the width of the page to the window
    pdftitle={My title},    % title
    pdfauthor={Author},     % author
    pdfsubject={Subject},   % subject of the document
    pdfcreator={Creator},   % creator of the document
    pdfproducer={Producer}, % producer of the document
    pdfkeywords={keyword1, key2, key3}, % list of keywords
    pdfnewwindow=true,      % links in new PDF window
    colorlinks=true,       % false: boxed links; true: colored links
    linkcolor=blue,          % color of internal links (change box color with linkbordercolor)
    citecolor=blue,        % color of links to bibliography
    filecolor=magenta,      % color of file links
    urlcolor=cyan           % color of external links
}

\newif\ifcomments
\commentstrue   %<<< Uncomment as required
\usepackage{subfiles}

\usepackage [english]{babel}
\usepackage [autostyle, english = american]{csquotes}
\MakeOuterQuote{"}

\author{Tonna Emenuga\thanks{Harvard University. Email: \protect\url{tonnaemenuga@college.harvard.edu}}}
\title{Filtering Down to Size: A Theory of Consideration\thanks{I am grateful to Tomasz Strzalecki for his guidance and mentorship. I thank Jerry Green, Shengwu Li, David Laibson, Yannai Gonczarowski, Nathaniel Hendren, Jeff Miron, Matthew Rabin, Angie Acquatella, Sejal Aggarwal, Shani Cohen, Roberto Colarieti, Benny Goldman, Zoë Hitzig, Martin Koenen, Pierfrancesco Mei, Akash Nandi, Cassidy Shubatt, Chris Walker and numerous seminar and workshop participants in the Harvard Economics Department for helpful discussions. This work was supported by a grant from the Harvard College Research Program. All errors are my own.}}

\begin{document}

\maketitle

\abstract{The standard rational choice model describes individuals as making choices by selecting the best option from a menu. A wealth of evidence instead suggests that individuals often filter menus into smaller sets — consideration sets — from which choices are then made. I provide a theoretical foundation for this phenomenon, developing a formal language of axioms to characterize how consideration sets are formed from menus. I posit that consideration filters — mappings that translate a menu into one of its subsets — capture this process, and I introduce several properties that consideration filters can have. I then extend this core model to provide linkages with the sequential choice and rational attention literatures. Finally, I explore whether utility representation is feasible under this consideration model, conjecturing necessary and sufficient conditions for consideration-mediated choices to be rationalizable.}

\newpage

\section{Introduction}

Economists' ability to deduce individual preferences from observed choices informs much of microeconomic theory. In particular, the foundational concept of revealed preference asserts that an individual's choice of an item (an alternative) from a set of options (a menu) is reflective of their underlying preference, allowing economists to determine preferences simply from a summary of choices made from various menus. For example, if when presented with two cars of equal cost\footnote{Throughout, I will assume all alternatives within a menu are affordable; this is a standard assumption in the literature.} — one red and one gray — a consumer purchases the red car, revealed preference tells us that the they must prefer the red car over the gray car.

In this standard model, individuals observe menus, analyze all available options, and make choices that are most consistent with their tastes. This model, which forms the foundation of rational choice theory as well as applied analysis across a variety of subfields, relies on an assumption referred to as "full consideration." This means that, when analyzing a menu, a decision-making individual considers \textit{every} item available before making a choice. In the car choice example, this entails assuming the individual considered the gray car, or was at least was aware of its presence. 

%As such, the standard model tells us that the individual in question will analyze all the available options and then pick the alternative that is most consistent with their tastes. 

%the standard model argues, is simply a matter of . Individuals are said to observe menus, analyze all available options, and simply pick the item alternative that is most consistent with their tastes. Such a paradigm - that choices and preferences are linked - relies on the assumption of every item in the menu was analyzed by the decision-maker. For ,xample, a consumer.

\par Contrary to this assumption, a wealth of empirical evidence demonstrates that, often, individuals will only consider a \textit{subset} of a menu before making a choice. In many cases, the entire menu is not fully examined: only the few alternatives which come to mind are fully considered by the individual. This is known as \textit{limited} consideration. Rather than making a choice directly from a menu, individuals may filter menus into \textit{consideration sets}, which are smaller groupings of alternatives from which choices are eventually made.

This filtered decision process creates some challenges for the classic revealed preference approach. For example, if an individual is presented with a menu consisting of alternatives $\{x, y, z\}$ and chooses $y$, can one state, as usual, that the individual necessarily prefers $y$ over $x$ and $z$? In the case of a filtered process, in which limited consideration holds, one cannot make this claim. Alternatives $x$ and $z$ may not have been in the individual's consideration set — $x$ and $z$ were not examined — and thus the preference relation of $y$ to $x$ and $z$ remains unknown.

\par Limited consideration also jeopardizes the ability attain utility representation of individual preferences. As is well known,\footnote{This is the utility representation theorem.} the ability to construct utility functions corresponding to observed choices relies on preferences being both complete\footnote{Preferences are complete if there is a well-defined relation between any two alternatives in a menu. Given two options, an individual with complete preferences will weakly prefer one of them, otherwise they are indifferent.} and transitive.\footnote{Preferences are transitive if $x \succsim y$ and $y \succsim z$ implies $x \succsim z$.} In the case of limited consideration, completeness is most clearly in question, and transitivity can fail as well.\footnote{\cite{masatlioglu_revealed_2012} provide several examples.} To better understand these issues, I provide in this paper a general model that can form the theoretical basis of work aimed at reconciling the standard model with the challenges imposed by limited consideration.

In doing so, I develop a formal language of axioms to characterize how individuals may not make choices from menus, but rather from consideration sets. I imagine that individuals observe menus, filter them into smaller menus, and then make rational choices from these smaller menus which are known as consideration sets. I posit that the process by which individuals go from menus to consideration sets is mediated by consideration \textit{filters},\footnote{Also referred to as consideration set mappings in the literature.} which are mappings that translate a given menu into one of its subsets.

Such a model has been developed in the literature. The idea that only a subset of available options are considered dates back as far as the \cite{simon1955behavioral} model of satisficing and optimal stopping, whereby an individual browses options only up until an acceptable one is found; at that point, search ceases. \cite{masatlioglu_revealed_2012} and \cite{masatlioglu2015completing} develop models to capture the process of limited consideration, focusing on the ability to infer consideration sets from observed choices under limited consideration.

\par On a normative level \cite{cherepanov2013rationalization} present a model in which individuals only consider alternatives that can be rationalized. \cite{ridout2021choosing} axiomatizes the decision-making process of an individual "choosing for the right reasons," modeling a decision maker who only makes choices that can be justified to others.

Such axiomatic formulations forms a solid theoretical grounding to make sense of much of the applied work on consideration. In particular, \cite{erdem1996decision}, \cite{hauser1990evaluation}, and \cite{roberts1991development} test structural models to describe the formation of consumers' consideration sets over goods. More recently, \cite{abaluck2016evolving} apply limited consideration to choices over healthcare plans, and \cite{abaluck2021consumers} develop a structural demand model based on limited consideration. 

\par I contribute to the literature by uniting much of the above work into a general framework for understanding limited consideration. I begin by formally outlining the main feature of the model: individuals' choice processes exhibit two mappings, one from menus to consideration sets and another from consideration sets to eventual choices. The timing of limited consideration features an individual observing a menu, considering some subset of the available alternatives (the consideration set), and then making a choice from said consideration set.

\par I model consideration sets as generated by consideration filters, functions that map the set of menus into subsets of themselves, thereby capturing the process of "filtering out" certain alternatives according to some heuristic. A consideration heuristic is simply a rule that determines which alternatives in a given menu are in the consideration set. For example, an individual intending to select a banana from a grocery store is unlikely to examine every banana in the produce section; rather, they may simply consider those bananas which are in the front row. In this case, the menu is the set of all bananas, the consideration set is the front row, and, importantly, the consideration \textit{filter} is the guiding heuristic "only look at bananas in the front row." Consideration heuristics may be intentional, in the case of the individual looking to purchase a banana, or subconscious, in the case of an individual aiming to select a fruit of any kind, and failing to see that there are apples, which they might very well prefer, in the next aisle. 

The list of potential consideration heuristics is clearly enormous, at least as large as the number of decision-making rules and behavioral biases that one could model an individual as having. I outline several generic qualities that such heuristics, or consideration filter properties, may have as well a the relationships that these properties have with one another. 

One of these properties, which I call Independence of Others (IO), is the focus of many of the exercises contained in the proceeding sections and in the paper as a whole. A consideration filter is IO if and only if alternatives in a menu are either always considered when available, or never considered at all. The sense in which this makes alternatives "independent" of each other is clear: the presence, or lack thereof, of other alternatives in a menu has no bearing on whether a particular alternative is in the final consideration set. IO closely approximates the rational model, insofar as IO filters generate choices structures that cannot cycle\footnote{Choices cycle if $x$ is chosen over $y$ and $x$ is chosen over $y$, yet $z$ is chosen over $x$, violating transitivity.} and hence can be represented by the utility function I derive in Section \ref{sec:ur}. I also present other potential filter properties corresponding to different consideration heuristics.

\par I then extend the basic model of consideration to account for the use of filters in a sequential fashion. The literature on sequential consideration begins with the \cite{tversky1972elimination} model of elimination by aspect, whereby one sequentially removes alternatives from the consideration set based on particular qualities — aspects — that these alternatives may or may not have. \cite{manzini2007sequentially} model a decision-making process whereby an agent eliminates inferior alternatives sequentially, applying a complete and transitive preference profile to the choice set until only one alternative remains — the final choice. \cite{apesteguia2013choice} take a game-theoretic approach, using game trees to characterize which sequential choice processes of the above sort are rationalizable. I propose a more general model, nesting the above models into a general framework for understanding multi-step consideration.

I then develop a second model extension, constructing a "rational attention"-style analog of the consideration model. This allows me to model preferences that decision-makers may have over consideration filters themselves, rather than simply over goods. In principle, there are may be a number of different consideration heuristics that an individual may employ, and the ultimate choice in large part rests upon which of these rules is applied to the menu they are presented with. I model an individuals who chooses which filter to apply to a given menu, weighing two competing forces: the benefit of a larger consideration set (a larger choice set may raise the chance of a particularly good alternative being in the menu) and the cost of examining many items (it takes time and effort to sift through many options). I derive two boundary conditions that give a flavor for how subsequent work can unite the process of consideration with the behavioral reality of limited attention.

\par Finally, I address the ability to rationalize choices that are made under limited consideration. The key threat to rationality, and utility representation, is completeness of preferences, which is clearly not the case when only a subset of available alternatives are considered. \cite{caplin2011search} discuss this challenge in a search-model setting. I provide a strong condition, IO, under which consideration-mediated choices can be represented by a utility function. Relatedly, I also provide a modification to the Weak Axiom of Revealed Preference (WARP) that matches the consideration setting, following in the vein of \cite{lleras2017more}. Overall, this paper outlines a general framework for understanding the phenomenon of limited consideration, extends the basic model in novel ways, and discusses the implications of limited consideration for utility representation and rational choice theory more broadly.

\par The paper proceeds as follows. Section \ref{sec:general} outlines the basic model of consideration filters and resultant consideration sets, proving a language through which the ideas of the papers can be discussed. Section \ref{sec:props} provides a number of qualities which consideration filters may have, in particular IO. Sections \ref{sec:seq} and \ref{sec:prefs} represent extensions to the base model: Sections \ref{sec:seq} allows for the use of more than one filter, and Section \ref{sec:prefs} models individuals as having preferences over filters. Section \ref{sec:ur} discusses the potential for utility representation of consideration-mediated choices. Section \ref{sec:fd} discusses the limitations of my results and future directions that the literature can take. Section \ref{sec:conc} concludes. An appendix$^{\ref{sec:app}}$, containing proofs of results, is also included.

\newpage

\section{General Theory of Consideration}\label{sec:general}

This section outlines the overall theory of limited consideration. I introduce the key players — consideration sets and consideration filters — while providing the surrounding definitions and axioms to close the model.

\subsection{Overview of Limited Consideration}

The basic theory of limited consideration is that an individual will not pay attention to every option they are presented with before making a choice. The simple translation of menus to choices is inadequate to capture a key nuance of human decision-making, namely that attention tends to be "costly" and thus individuals consider only a small set number of alternatives in the choice process. Between menus and choices, there exists an intermediate step known as \textit{consideration}, which defines the smaller set of alternatives that an individual will examine and ultimately choose from. 

Rather than making choices from menus, individuals make choices from consideration sets, which are nested within the original menus. Figure \ref{fig} makes clear the relationships between menus, consideration sets, and choices.
\vspace{0.2cm}
\begin{center}
    \includegraphics[scale=0.40]{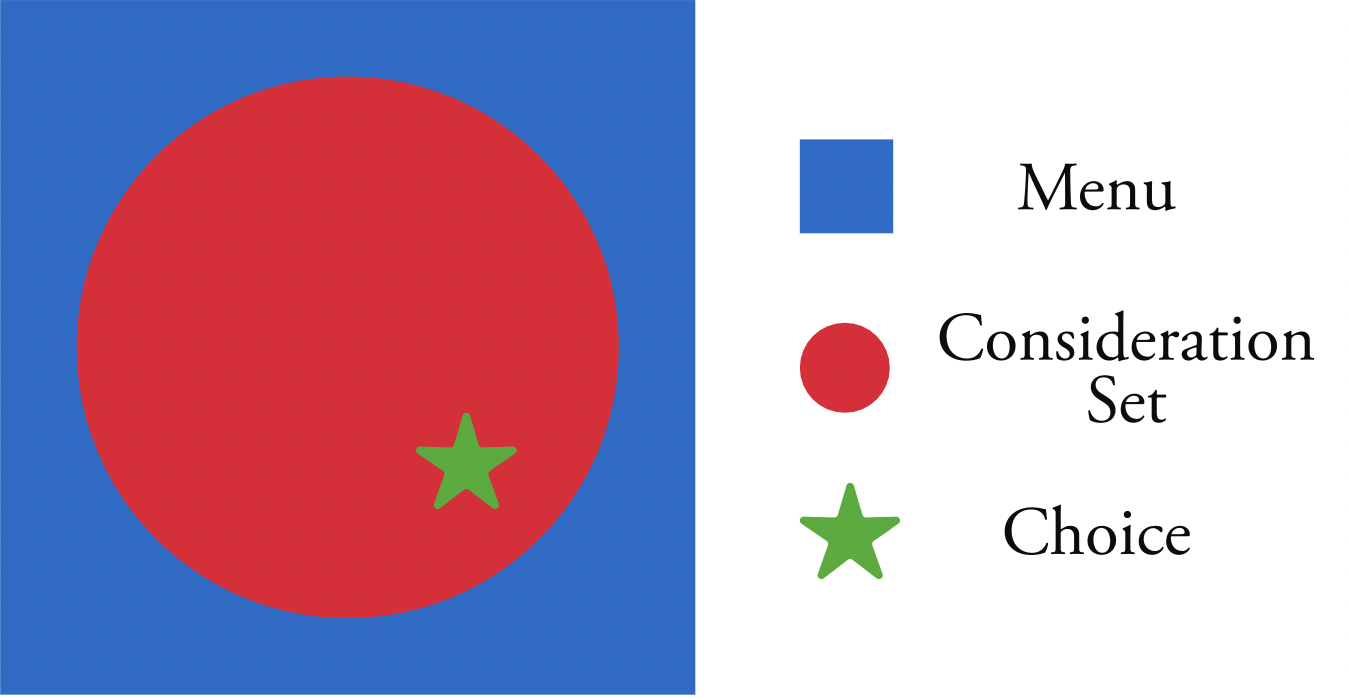}
\end{center}
\begin{center}
    Figure 2.1: Menu, Consideration Set, and Choice \label{fig}
\end{center}

\vspace{0.2cm}

I now provide the formal definitions and axioms to set up the environment.

\subsection{Definitions and Axioms}

Equipped with an intuitive notion of the idea of consideration sets, I now define the mathematical environment in more concrete terms. This section will use concepts that are quite familiar to the decision theorist and discrete mathematician alike.

\subsubsection{Set Notation}

I use basic set-theoretic concepts to define the terms used in the process consideration-mediated choice.

\begin{definition}
    The set of alternatives is X.
\end{definition}

The set of alternatives represents the total set of alternatives which may be presented to an individual. For an individual browsing cars at a dealership, $X$ is the total set of cars in the lot. The set of cars consists of individual cars, which are its constituent alternatives:

%For example, an individual browsing cars at a dealership wishes, in principle, to choose some car from the total set of cars available. This total set of cars at the dealership is $X$. Limited consideration, and the fact that most menus will be smaller than the total set of options, imply that the individual will rarely in actuality face the entire set $X$ - the menu that the individual observes, which I will define, is likely some subset of this set $X$.Extending the car example, each individual car will count as an alternative, which we denote as $x_i$: 

\begin{definition}
    The constituent alternatives within $X$ are known as $x_i$, where $x_i$ $\in X$ for all $i$.
\end{definition}

The individual's final choice will be some alternative $x_i$ from the universe $X$. I henceforth abuse notation by omitting the subscript $i$ to refer to alternatives by $x$. As discussed, individuals must first observe a menu — some subset of the available alternatives. 

%Hearkening back to Figure 1, the set of alternatives $X$ contains the blue region that is the individual is presented with. For simplicity, I omit the set $X$ from the graphic.

\begin{definition}
    The set of menus is M(X).
\end{definition}

The set of menus constitutes every combination of alternatives $\{x_1, x_2, ..., x_n \}$ that the individual may be presented with.

\begin{axiom}
    There exist $2^{|X|}$ menus.
\end{axiom}

Following from the definition of the power set, the set of menus $M(X)$ will necessarily contain $2^{|X|}$ menus, where $|X|$ is the cardinality of $X$, the number of alternatives contained in it. This includes both the full set, containing every alternative, as well as the null set $\emptyset$.

\begin{definition}
    Generic menus are denoted $A$ and $B$; $A \subseteq B \in M(X)$.
\end{definition}

I define two generic menus $A$ and $B$, which will be used in proceeding sections when discussing properties of the model. $A$ is a subset of $B$ — every alternative in $A$ also finds itself in $B$. I later sections, I will at times make use of other generic menus, such as $Q$; in each instance, take $Q$ to be some generic menu with the same properties as $A$.

\subsubsection{Mappings and Consideration Sets}

I now define consideration filters and consideration sets. The consideration model assumes that individuals observe menus, filter them into consideration sets, and then make choices. The first step in the process is the mapping from menus to consideration sets. Formally, a consideration filter is a mapping $\Gamma: M(X) \mapsto M(X)$ that takes the set of menus and returns corresponding subsets. That is, a consideration filter takes a menu and returns one of its subsets. For the menu $A$, the filter returns $\Gamma(A)$, where $\Gamma(A) \subseteq A$ for all menus $A \in M(X)$. Recall the generic menus $A$ and $B$, keeping in mind that $A$ is a subset of $B$. A consideration filter, in the most general sense, takes $B$ and returns some $A$:

\begin{definition}
    $\Gamma: B \mapsto A$ is a consideration filter.
\end{definition}

Any function that takes a set and returns one its subsets qualifies as a consideration filter, capturing the phenomenon of the limited consideration that motivates the literature. Naturally, such filters can take various forms. For example, \cite{masatlioglu_revealed_2012} restrict their analysis to those filters classified as "attention filters," while \cite{lleras2017more} study the case of "choice overload."

\begin{definition}
    $\Gamma(B)$ is the consideration set of menu $B$.
\end{definition}

The result of the consideration filter, applied to a menu, is the consideration \textit{set}, represented by the red region of Figure \ref{fig}. Choices, ultimately, are made from consideration sets rather than menus.

\begin{axiom}
    $c(B) \in \Gamma(B)$
\end{axiom}

The above axiom simply establishes that the choice from a menu must also be in that menu's consideration set. In essence, choices are made from consideration sets, not menus. This provides a starting point for any attempts to identify consideration sets from observed choice data. If an alternative is chosen, one can state with certainty that said alternative must have been considered, and thus finds itself within the consideration set.\footnote{Note that, if an alternative is \textit{not} chosen, one cannot make any inference on whether it was considered, without making additional assumptions on the consideration process.} In order to determine the structure consideration sets beyond the chosen alternative, assumptions must be made on the process of consideration, necessitating some language for describing axioms of consideration filters. The following section provides such a framework.

\section{Properties of Consideration Filters}\label{sec:props}

This section introduces various properties that consideration filters may have, providing a language through which consideration filters and consideration sets can be described. Section \ref{sec:propsenrivo} re-states the mathematical preliminaries, outlining a rigorous vocabulary for describing the environment of interest. Section \ref{sec:propslist} lists properties that consideration filters may have, providing examples, descriptions, and formal results regarding their interrelations when necessary.

\subsection{Environment}\label{sec:propsenrivo}

I reformulate the mathematical environment, using basic notions of sets and set relationships. I begin with an individual, who goes through a two-step process to make a choice: the individual is presented with a menu of items, filters it down to a weakly smaller consideration set, and then makes a final choice. In this section, I focus on the first portion of the process: that is, the transformation of menus into consideration sets.

\par There is a set of alternatives, $X$, which captures the universe of goods that the individual may potentially observe. The space $X$ is non-empty, instead having constituent elements $x \in X$, each of which is a particular alternative which could be chosen. However, individuals do not ordinarily observe the entire set of alternatives $X$: while this is possible, individuals more commonly are presented with \textit{menus}, smaller collections of alternatives which are drawn from the larger menu. 
%For example, upon entering a grocery store, one observes a set of fruits, which are in principle only a subset of the entire set of fruits which exist.

\par Formally, $M(X)$ denotes the set of menus which can be derived from $X$, each of which is a subset of X: for a menu $A$, $A \in M(X) \subseteq X$. There are $2^X$ potential menus, including the full set $X$ and the null set ${\emptyset}$.

\par I refer to $A$ as being a generic menu in $M(X)$ and also make use of a generic menu $B$ representing a menu that constitutes a superset of menu $A$: $A \subseteq B$.

\par Individuals observe a menu and consider a subset of its alternatives. To describe this, I introduce a mapping, a consideration filter, which takes an observed menu $A \in M(X)$ and produces another menu $\Gamma(A)$. Because an individual can only consider alternatives which are in the original menu, the consideration set is a subset of the original menu: $\Gamma(A) \subseteq A$. The consideration filter is, then:

\begin{equation}
    \Gamma: M(X) \mapsto M(X)
\end{equation} 

\par The following section, a listing of properties of consideration filters, puts more structure on the nature of consideration filters.

\subsection{List of Properties}\label{sec:propslist}

The following list of properties describes various features that a consideration filter may have. A given consideration filter may satisfy some number of these properties, all of them, or none at all.. Note that, while these properties are distinct, they are not necessarily mutually exclusive, nor are they, as a rule, concomitant.

\par The first two properties represent adaptations of well-known axioms in rational choice theory, translated here to axiomatize consideration filters rather than choice functions. In the standard choice literature, axioms are frequently presented in the following style:
\begin{example}\textsc{A Generic Choice Axiom}
    If $x$ is chosen from menu $A$, it must be chosen from menu $B$.
\end{example}

The weak axiom, for example, takes this basic form. The idea behind this approach is that, by making assumptions on the details of the decision process, one can make claims about which alternatives from menu are chosen from another menu, simply from taking note of the latter menu's structure.

In this section, I adapt this familiar axiomatic style to accommodate \textit{consideration} rather than the final choice. That is, I present a group of axioms that, under certain conditions, give clarity on which alternatives from a given menu are in the consideration set. In doing so, I extend the rigour of axiomatic choice theory into the realm of consideration, providing the literature with a set of sample axioms to form a basis for further exploration and applied work. 

Each axiom I present represents a property that a consideration filter may have, outlining a condition under which the consideration set of a menu can, at least in part, be identified. I pair each axiom with an accompanying example. Throughout these examples, I make the following assumptions:

\begin{itemize}
    \item The set of alternatives is $X: \{1, 2, 3, 4, 5, 6, 7, 8, 9\}$.
    \item The set of menus $M(X)$ contains $2^9=512$ menus.
    \item $A$ is a generic menu; $A: \{1, 2, 3\}$
    \item $B$ is a generic menu; $B: \{1, 2, 3, 4, 5\}$
\end{itemize}

These assumptions preserve generality and are made for the purpose of concretizing each axiom. I also introduce generic menus other than $A$ and $B$ is specific cases. I begin with Sen's $\alpha$:

\begin{definition}\textsc{Sen's $\alpha$}. $\Gamma$ satisfies Sen's $\alpha$ if $x$ $\in$ $\Gamma (B)$ and $x$ $\in$ A $\subseteq$ B implies $x$ $\in$ $\Gamma (A).$
\end{definition}

This is the well-known Sen's $\alpha$\footnote{Also known as Chernoff's condition.} axiom, adapted to the consideration setting. In short, Sen's $\alpha$ asserts that, whenever $x$ is considered from a set $B$, $x$ will also be considered from all of $B$'s subsets in which it is present. Accordingly, I call this a "large-to-small" condition, guaranteeing that any alternative considered from a larger set will also be considered in its subsets.

\begin{example}\textsc{Sen's $\alpha$ example}
\end{example}
Suppose $\Gamma$ satisfies Sen's $\alpha$. Further suppose there is a menu $B = \{1, 2, 3, 4, 5\}$ with consideration set $\Gamma(B) = \{2, 3\}$. If $A:\{1, 2, 3\}$, then $2, 3 \in \Gamma(A)$.

This example shows how Sen's $\alpha$, when assumed to hold with respect to a given consideration filter $\Gamma$, allows an observer to determine which alternatives are in the consideration set of another menu. One salient point to note is that the consideration set of $A$, in this case, \textit{is not limited} to 2 and 3. It is possible that 1 is also in the consideration set of $A$. However, given Sen's $\alpha$, we can only say \textit{for sure} that 2 and 3 are in the consideration set of $A$. These axioms provide a lower bound on the alternatives within a menu's consideration set.

Sen's $\beta$ is another familiar result, here reformulated to match the consideration filter environment.

\begin{definition}\textsc{Sen's $\beta$}.
$\Gamma$ satisfies Sen's $\beta$ if $x_1, x_2 \in A \subseteq B$ and $x_2 \in \Gamma(B)$ implies $x_1 \in \Gamma(B)$.
\end{definition}

Sen's $\beta$ is another classical axiom of decision theory that I analogize to the consideration setting. The axiom asserts what one might call the "take-with-me" property: if two alternatives are considered from a menu, and then available in a larger menu, then one cannot be considered without the other\footnote{Of course, this abstracts away from externally-imposed limits on the size of consideration sets. I address this in Section \ref{sec:prefs}.}.

\begin{example}\textsc{Sen's $\beta$ example}
\end{example}

Suppose $\Gamma$ satisfies Sen's $\beta$. Further suppose there is a menu $A = \{1, 2, 3\}$ with consideration set $\Gamma(A) = \{2, 3\}$. If $2 \in \Gamma(B)$, where $B \supseteq A$, then $3 \in \Gamma(B)$.

\vspace{0.2cm}

The next property is a novel axiom I present as a counterpart to Sen's $\alpha$.

\begin{definition}\textsc{Condition $\tau$}.
$\Gamma$ satisfies Condition $\tau$ if x $\in$ $\Gamma (A)$ and A $\subseteq$ B implies x $\in$ $\Gamma (B)$.
\end{definition}

Condition $\tau$ "reverses" Sen's $\alpha$, considering a "small-to-large case" rather than a large-to-small. Condition $\tau$ states that if an alternative $x$ is in the consideration set of a menu $A$, then it is in the consideration set of every superset of $A$ in which it is present. This gives a characterization of the preservation of consideration sets under "expansion". When Condition $\tau$ holds, no alternatives are dropped from consideration during this expansion.

Note that filters satisfying this Condition $\tau$ preclude the possibility of choice overload, the "more is less" phenomenon described by \cite{lleras2017more}. The concept of choice overload posits that adding items to a menu may result in previously-considered alternatives ceasing to be a part of the new consideration set, mirroring the well-known literature of "overwhelmed" consumers losing track of alternatives and thus dropping goods from consideration when confronted with larger menus. Condition $\tau$ is a direct remedy for this, although it does not not apply realistically to every setting of interest. %I can show that Condition $\tau$ makes possible the identification of filters satisfying certain properties, in particular Independence of Others. Later sections expand on this.

\begin{example}\textsc{Condition $\tau$ example}
\end{example}
Suppose $\Gamma$ satisfies Condition $\tau$. Further suppose there is a menu $A = \{1, 2, 3\}$ with consideration set $\Gamma(A) = \{2, 3\}$. For menu $B \supseteq A$, where $B = \{1, 2, 3, 4, 5 \}$, $2, 3 \in \Gamma(B)$.

%While $A$ is our generic menu in $M(X)$, call $J$ a generic menu that is also in $M(X)$. We do not impose any conditions on $J$, namely with respect to whether it is a subset or superset of $A$.

\begin{definition}\textsc{Independence of Others}
$\Gamma$ satisfies Independence of Others (IO) if the following condition holds:
\begin{enumerate}
    \item \textit{x $\in$ $\Gamma (A)$ $\forall$ A $\in$ M(X) s.t. x $\in A$} \textbf{or}
    \item \textit{x $\notin$ $\Gamma (A)$ $\forall$ A $\in$ M(X)} 
\end{enumerate}
\end{definition}

Independence of Others, when true, states that every alternative $x \in X$ is either always considered whenever it is available (i.e. in the menu), otherwise it is never considered. This gives an intuition for the name: an alternative's inclusion in the consideration set is purely independent of the the presence of all other alternatives. There is no comparative process at this stage of the decision-making process. Naturally, IO corresponds to cases in which there exist certain properties of goods that are not subject to change, and individuals form consideration sets based on these unchangeable qualities. For example, an individual deciding which car to purchase at a dealership, but restricting their consideration to cars that travel at least 30 miles per gallon, chooses to consider cars in a way that is IO. Any car in the available set with at least 30 MPG is in the consideration set, \textit{independent} of whether any other car is in the set. Most consideration heuristics are not IO, particularly those which make consideration dependent on comparison or position in some order. For example, the heuristic "when shopping for fruits, only consider bananas in the front row" is not IO because moving a banana's position affects whether or not it is considered. IO consideration heuristics can, more so than other rules, generate null consideration sets in the event that no alternative in the menu satisfies the criterion of interest.

IO matches the standard rational choice model in the sense that revealed preference analysis is always possible over alternatives that are ever chosen. Recall that the main threat to revealed preference is incomplete preferences caused by lack of consideration. IO largely closes this hole by guaranteeing that any alternative that chosen from any menu is considered across all menus in which it is available. This gives greater scope to infer preferences from observed choices. 

\par Of course, IO is not realistic in every setting, and this exposes a weakness in the standard rational choice model. Any preference ordering taken directly from observed choices implicitly assumes full consideration, which IO approximates. Thus, this presents a challenge for economists who maintain the usefulness of the standard revealed preference framework while naturally disputing the accuracy of IO in describing real-world decision-making.

\begin{example}\textsc{IO example}
\end{example}
Suppose $\Gamma$ satisfies IO. Further suppose there is a menu $A = \{1, 2, 3\}$ with consideration set $\Gamma(A) = \{2, 3\}$. If we have menus $B = \{1, 2, 3, 4, 5 \}$, $Q = \{1, 4, 7\}$, and $S \{1, 6, 8\}$, $1 \in \Gamma(B)$, then $1 \in \Gamma(Q)$, and $1 \in \Gamma(S)$. 
\vspace{0.2cm}

I now introduce a result regarding the relationship between IO and two preceding properties.

\begin{theorem}\textsc{(Sen's $\alpha$ and Condition $\tau$ $\Leftrightarrow$ IO)} $\Gamma$ satisfies Sen's $\alpha$ and Condition $\tau$ if and only if it satisfies IO. 
\end{theorem}

\begin{proof}
See appendix.
\end{proof}

Intuitively, Sen's $\alpha$ and Condition $\tau$ jointly provide both a "large to small" as well as a "small to large" condition on filter $\Gamma$. The proof strategy relies on the singleton set $\{x\}$ being a subset of any menu $A$ that includes $x$. If $x$ is in the consideration set of $A$, $\Gamma(A)$, one can guarantee consideration of $x$ in any menu that includes by going "down and up," a technique I demonstrate in the proof.

The next property, Dynamic Independence of Others (DIO) \footnote{Sejal Aggarwal coined the name of this property, helping me replace a previous term that was not as clear.}, extends IO, and the process of consideration, into a dynamic setting. DIO states that, when alternatives in a menu are filtered in some order, the order does not affect the final consideration set.

To characterize DIO formally, fix the elements of menu $A \in$ M(X) so as to consider every possible ordering of its elements. For example, the $A = \{1,2\}$ can equivalently be represented as $A = \{2,1\}$. For a menu $A$ with $|A|$ elements, there exist $|A|!$ possible orderings. Call the first ordering $A_1$ and the second $A_2$, giving us a general notation where $A_n$ denotes the $nth$ ordering of menu $A$. For a menu $A$, imagine an individual is endowed with $|A|$ units of time, during each of which a singular element is considered. That is, in period 1, the individual considers the first element, then in period 2 the second element... until the final element is considered in period $|A|$. 

Dynamic Independence of Others asserts that the final consideration set will be equal among each such ordering.

\begin{definition}\textsc{Dynamic Independence of Others}
$\Gamma$ satisfies Dynamic Independence of Others (DIO) if $\Gamma(A_1) = \Gamma(A_2) = \cdot \cdot \cdot =  \Gamma(A_n) $ where n = $|A|!$.
\end{definition}

DIO rules out the possibility behavioral biases that make choices a function of the manner in which alternatives are presented, rather than simply on the combination of alternatives which constitute the observed menu. For this reason, DIO represents a very strong condition on filtering in dynamic environments. 

\par Consideration heuristics satisfying DIO are incompatible with satisficing behavior of the sort described by \cite{simon1955behavioral}. In Simon's model, consumers view alternatives $x \in X$ in a set order, and make a choice once some alternative meeting quality threshold is observed. Clearly, order matters here. For example, in a setting with 10 alternatives where 5 are "satisfying" but only 3 are to be considered, the 3 out of 5 satisfying alternatives which come first will be considered, while 2 "miss out." DIO precludes this from occurring. Therefore, it is not applicable to satisficing scenarios and perhaps a wide array of models in which one assumes a similar "first-mover" effect. However, DIO gives the standard consideration filter model a benchmark against which such dynamic behavioral frameworks can be tested.
%and may only consider the first $n$ alternatives meeting a certain threshold, forming the consideration set when a sufficient number of satisfactory alternatives have been considered. 

\begin{example}\textsc{DIO example}
\end{example}
Suppose $\Gamma$ satisfies DIO and consideration is made in a ordered fashion. Further suppose there are menus $A_1 = \{1, 2, 3\}$, $A_2 = \{1, 3, 2\}$, and $A_3 = \{3, 1, 2\}$. $\Gamma(A_1) = \Gamma(A_2) = \Gamma(A_3)$.

\vspace{0.2cm}
  
The next property, Constant Number, makes use of the notion of cardinality, the number of elements in a set. As stated earlier, call $|M(X)|$ the number of elements (cardinality) of a menu M(X), by virtue of which $|\Gamma(A)|$ is the number of elements considered from a menu $A$. For example, if $A = \{1, 2\}$, then $|A| = 2$.

\begin{definition}\textsc{Constant Number (CN)} 
$\Gamma$ satisfies Constant Number if $|\Gamma(A)| = n$ for all $A \in$ such that $|A| \geq n$.
\end{definition}

A constant number consideration heuristic always considers the same number of alternatives from a menu, provided that the menu has that number of alternatives available. This requires one to fix some $n \in \mathbb{W}$.\footnote{$\mathbb{W}$ is the set of natural numbers and 0.}

\begin{example}\textsc{Constant Number example}
\end{example}
Suppose $\Gamma$ satisfies CN, with $n = 2$. Further suppose there is a menu $A = \{1, 2, 3\}$. $Gamma(A)$ must be $\{1, 2\}$, $\{1, 3\}$, or $\{2, 3\}$.

\subsection{Summary}

This section restated the formal language of consideration sets and consideration filters, allowing me to define various properties which consideration filters can potentially satisfy. These properties are neither mutually exclusive as a rule nor necessarily concomitant. The first two properties, Sen's $\alpha$ and Sen's $\beta$, are adaptations of well-known 
 choice axioms, herein translated to match the setting of interest: mappings from menus to consideration sets, rather than the standard mapping from menus to choices. I then introduce several novel properties that consideration filters may have: Condition $\tau$ Independence of Others (IO), Dynamic Independence of Others (DIO), and Constant Number.

 \par Condition $\tau$ reverses Sen's $\alpha$ to preclude the possibility of the choice overload phenomenon. IO is a very strong condition, mandating that an alternative is either always considered when available, or never considered. DIO extends IO to a dynamic setting. As will be shown is later sections, IO matches the rational model and, while perhaps unrealistic, is necessary for utility representation under the standard model. Constant Number makes all consideration sets equal in cardinality.

 \par In the next section, I extend the basic consideration model to allow for the use of more than one consideration filter on a given menu.

%CONCLUSION: These three conditions give us greater clarity about the relationships between the properties discussed in \textbf{Section 3}. In particular, the features of IO and SIO are made more salient. \textbf{Section 3} now considers an extension to the base model: the application of more than one filter to the same menu before choices are made.

\section{Extension 1: Sequential Consideration}\label{sec:seq}

Imagine an individual walks into a store looking for a new pair of shoes. In narrowing down options, they focuses their attention on shoes in the first aisle. They then only consider shoes of size 10, further narrowing down their options. How might one model this process? As it stands, the current consideration model can only account for one consideration filter, however this example requires two: one from all shoes to those in the first aisle, and then another from those in the first aisle to those which are \textit{also} size 10. A extension on the basic framework is needed to account for this.

\par In this section, I propose and develop a model of sequential consideration. While the idea of consideration sets has been explored in the literature, \textit{two-step}, or even \textit{n-step} (with more than 2 rounds of narrowing down), has not been formally explored. Multiple rounds of consideration may better match real-world settings, by covering scenarios in which individuals narrow down large sets based on multiple criteria.

Recall that, in the original case, we have the following mapping from menus to consideration sets to eventual choices:

\begin{equation}
    A \mapsto \Gamma(A) \mapsto c(\Gamma(A))
\end{equation}

Whereas there could exist more than one consideration filter:

\begin{equation}
    M(X) \mapsto \Gamma_{1}(A) \mapsto \Gamma_{2}(A) \mapsto c(\Gamma_{12}(A))
\end{equation}

\par One might imagine Figure \ref{fig}, in this case, having series of circles \textit{within} the original one, each of which represents a new downsizing of the consideration set. I provide a formal model for this, first by constructing a space of filters that contains any of the many consideration filters an individual may use. These consideration filters may satisfy different properties. For example, one filter may be IO and another may be CN. Individuals can apply any number of filters to a given menu, narrowing it down to a final consideration set in multiple steps.

\par This framework has precedent in the literature. \cite{tversky1972elimination} introduced the well-known process of elimination by aspect. In his model, menus contain alternatives, each which has or does not have some aspect — a desirable features of goods. Individuals make choices in a multi-step process, where each step involves eliminating all alternatives which do not have a particular aspect. The process continues until only one alternative is left, which becomes the final choice. \cite{manzini2007sequentially} imagine an individual using multiple "rationales" — complete and transitive preference relations — to a given menu, applying such rationale in a fixed order to arrive at a choice. The authors then evaluate which sorts of choice functions are consistent with the unique alternative selected by such a process. \cite{apesteguia2013choice} provide a taxonomy for the sort of model specified by \cite{manzini2007sequentially}, using game trees to formalize the idea of sequential rationalizability.

\par I propose a more general model, nesting the above models into a general framework for understanding consideration in multiple steps. For example, elimination by aspects can be modeled as applying a set of IO filters to a given menu, given that each "aspect" represents an immutable characteristic of an alternative. My model differs from that of \cite{manzini2007sequentially} in that I do not make use of rational preferences until \textit{after} the consideration set has been formed. I take a different approach from \cite{apesteguia2013choice}, grounding my analysis in the standard choice framework rather than the style of game-theoretic decision trees. 

\par In addition, I introduce a property, commutativity, borrowing the term from algebra to characterize relations between two or more filters. I say that two or more filters are commutative if their successive application to a menu produces the same consideration set, regardless of the order in which the filters are applied. Commutativity, as a concept, has applications to any decision setting in which choices may be contingent on the manner in which information is presented. I show that any number of IO filters are always commutative, proving two results.

% To this end, this section outlines a formal way to think about this phenomenon: sequential consideration. I present a formal language for discussing such consideration heuristics, presenting a vocabulary for doing so and leading into 2 key results.

\subsection{Definitions}

I begin by defining the space of filters, which contains the mass of consideration filters that can be applied to a menu.

\begin{definition}\textsc{Space of Filters}
There exists a space of filters $\gamma$, comprising constituent filters $\Gamma_i \in \gamma$. $\Gamma_{i}$ represents the ith filter.
\end{definition}

This defines the space of filters that are available to a decision maker, each of which narrows a menu according to whatever properties it has and the heuristic that is implicitly associated with it. Any filter that is applied to a menu comes from $\gamma$. The example above involves $\Gamma_1$ and $\Gamma_2$, two are consideration filters which come from $\gamma$ and are applied to menu $A \in M(X)$. There are multiple ways to represent two filters applied to a menu.

\begin{definition}\textsc{Representation of 2-Step Consideration}
If $\Gamma_{1}$, and then $\Gamma_{2}$, are applied to menu $A$, the resultant consideration set is $\Gamma_{2}(\Gamma_{1}(A)$ or $\Gamma_{12}$.
\end{definition}

Notice that the nested notation, $\Gamma_{2}(\Gamma_{1}(A)$, reads from right to left, whereas the reduced form $\Gamma_{12}$ reads from left to right.

The timing of 2-step consideration proceeds as follows: an individual observes a menu, applied a consideration filter to that menu to get a consideration set, and then applies \textit{another} filter the consideration set to end with a final consideration set. As such, consideration filters can be analogized to contracting mappings in dynamic programming: each one, applied in succession, takes the current menu and returns a smaller menu, which is still in the space of menus $M(X)$. Similar notation and intuitions hold for $n$-step consideration, in which any finite number of consideration filters within $\gamma$ are applied to a menu in succession:

\begin{definition}\textsc{Representation of \textit{N}-Step Consideration}
If $\Gamma_{1}$, $\Gamma_{2}$, ..., $\Gamma_{n}$ are applied to menu $A$, the resultant consideration set is $\Gamma_{n}(...(\Gamma_{2}(\Gamma_{1}(A))))$ or $\Gamma_{12...n}$.
\end{definition}

Here, more than one filter can be applied to a menu. The timing of this process is principle the same as that of 2-step consideration, albeit with a potentially large number of new, smaller considerations sets being formed with each application of a filter $\Gamma_i$ from $\gamma$. 

Equipped with notation to describe sequential consideration, I now introduce the notion of commutativity, extending the framework to detail the importance of the order in which a set of filters is applied.

\subsection{Commutative Filters}

When different filters are applied sequentially to a menu, does order matter? That is, can one apply them in any order and expect to get the same final consideration set? As a concrete example, imagine an individual asking a librarian for help selecting a new book to read. Alarmed by the incalculably-high number of books to choose from, they use two rules of thumb: they want to read fiction, and also want to read one of the first books that comes to the librarian's mind when asked. The individual will make both of these requests, but could either ask for fiction and \textit{then} for the first few books that come to mind, or ask for the first few books that come to mind, and then ask which of these books are fiction. This section addresses whether the consideration set from which they end up choosing the will be the same in both cases.

In order to answer these questions, I introduce a new concept: commutativity. I adapt the basic axiom of algebra here to describe cases in which consideration filters can be applied in any order to a menu, without the consideration set changing. Simply put, order does not matter. As one may expect, this is a rather strong condition. In particular, its viability will often depend on whether Independence of Others (IO) is satisfied by the filters in question. Before introducing results, I define commutativity in both the 2-step and $n$-step cases. Without loss of generality, $\Gamma_1$ and $\Gamma_2$ refer to two, distinct, arbitrary filters in $\gamma$.

\begin{definition}\textsc{2 Commutative Filters}
$\Gamma_{1}$ and $\Gamma_{2}$ are commutative if $\Gamma_{12}(A) = \Gamma_{21}(A)$ for all $A$. That is, $x \in \Gamma_{12}$ iff $x \in \Gamma_{21}$.
\end{definition}

Two filters are commutative if their application to a menu $A$ generates the same final consideration, \textit{regardless of the order in which they are applied.} This provides language to describe consideration sets that are invariant to the order of 2 filters that generated them. I extend this definition to the $n$-step case, which is a more demanding condition. 

%This gives us some notion of preservation which is invariant to ordering. 

\begin{definition}\textsc{\textit{N} Commutative Filters}
N filters $\Gamma_{1}$, $\Gamma_{2}$, ..., $\Gamma_{n}$ are commutative if $\Gamma_{12...n}(A)$ is invariant to permutations in the order of $\{1,2,...n\}$.
\end{definition}

Commutativity for $n$ filters works in the same way that it does for 2 filters. However this DIO requires checking every permutation, and the set of filter orderings can become very large. For example, for a set of 5 filters, there are 120 different orderings, each of which must be verified to determine if commutativity holds. These definitions allow to present the two main results of this section.

\subsection{Commutativity Results}

Commutativity necessarily makes certain demands on the properties that the applied consideration filters must have. I present two results, which show that IO filters are necessarily commutative, both in the 2-step and $n$-step cases. Recall the definition of IO:

\begin{re-definition}\textsc{Independence of Others}
A consideration filer $\Gamma$ satisfies Independence of Others (IO) if one of the following two conditions holds:
\begin{enumerate}
    \item \textit{x $\in$ $\Gamma (A)$ $\forall$ A $\in$ M(X) s.t. x $\in A$} \textbf{or}
    \item \textit{x $\notin$ $\Gamma (A)$ $\forall$ A $\in$ M(X)} 
\end{enumerate}
\end{re-definition}

IO maintains that each alternative is always considered when available, otherwise else it is never considered. I now introduce the first result, which requires IO for two filters to be commutative for any arbitrary menu.

\begin{theorem}\textsc{IO and Commutativity with 2 Filters}
%$\Gamma_{1}$ and $\Gamma_{2}$ are commutative for all $A \in M(X)$ if and only if $\Gamma_{1}$ and $\Gamma_{2}$ are IO.
\end{theorem}
$\Gamma_{1}$ and $\Gamma_{2}$ are commutative for all $A \in M(X)$ if and only if $\Gamma_{1}$ and $\Gamma_{2}$ are IO.
\begin{proof}
See appendix.
\end{proof}

I wish to place emphasis on the fact that, while IO is necessary for two filters to be commutative for \textit{any arbitrary menu}, two filters can be commutative for a given menu, while not being commutative for \textit{all} menus. A simple possible example is if the menu is the null set $\emptyset$. Any two filters are commutative for this menu, as the consideration set remains null, while, if they are not IO, order reversal could change the consideration set generated from a non-empty menu.

\begin{theorem}\textsc{IO and Commutativity with $N$ Filters}
%$\Gamma_{1}$, $\Gamma_{2}$, ..., $\Gamma_{n}$ are commutative for all $A \in M(X)$ if and only if $\Gamma_{1}$, $\Gamma_{2}$, ..., $\Gamma_{n}$ are IO.
\end{theorem}
$\Gamma_{1}$, $\Gamma_{2}$, ..., $\Gamma_{n}$ are commutative for all $A \in M(X)$ if and only if $\Gamma_{1}$, $\Gamma_{2}$, ..., $\Gamma_{n}$ are IO.

\begin{proof}
See appendix.
\end{proof}

This extends Theorem 2 to cases with potentially more then 2 filters, although the degenerate case of $n=2$ shows us that Theorem 3 nests Theorem 2, trivially. The proof idea, executed in the appendix, is to use the 2-step proof as a base case for induction.

\subsection{Summary}

This section introduces the idea of sequential consideration, the idea that multiple filters can be applied to a given menu, each of which entails a new contraction of the consideration set. Filters come from the larger space of filters $\gamma$, and may satisfy any number of properties. I then introduce the concept of commutativity, which specifies cases in which 2, or any countable number of filters generate the same consideration set regardless of the order in which they are applied. Commutativity will always hold among any number of IO filters, and can hold among non-IO filters in more limited cases. Generally, commutativity is a useful benchmark in thinking about consideration within settings in which information acquisition, and the decisions made from such information, are paramount. This allows decision theorists to specify when and under what conditions information use is unaffected by the order of its arrival. In this next section, I use the idea that there may exist multiple filters in a decision process to endogenize the choice of a filter within the rational-attention context.

% The next section takes a different spin on the original model of \textbf{Section 3}, aiming to define the notion of preferences with respect to filters.

\section{Extension 2: Preferences over Filters}\label{sec:prefs}

To this point, the concept of consideration filters has been well-developed, in particular an outline of their potential properties and interrelations. In this section, I model individuals as having preferences not simply over alternatives, but over \textit{filters} that affect the set of alternatives with which they are presented.

Such a formulation is well-suited to the nuances of individual decision making. Choice naturally involves rules of thumb — filters, as I model them — however, individuals may apply such rules selectively across settings. The best consideration heuristic for choosing a car will naturally differ from that which is optimal for shopping for bananas. I there endogenize the choice of filters, allowing individuals in my model to select which consideration filter to apply to a given menu. For simplicity, I assume in this section that individuals only choose one filter, abstracting away from the sequential setup of the last section. In practice, the two concepts are easy to combine; I herein wish to focus on the preference portion to make the concept clear. I place my model in the context of the rational attention literature, positing that there are two competing factors in choosing a filter: the greater optionality provided by a larger consideration set, and the costly mental strain associated with sifting through numerous options. I build a parsimonious model to capture the substance of this idea, while omitting certain details that risk over-complicating the setup.

\subsection{Environment and Details}

I now formally characterize the space in which the definitions, axioms, and results to follow shall operate. In Section \ref{sec:propsenrivo} I outlined the environment, particularly defining the relationship between the set of alternatives $X$ and its constituent menus $A \in M(X)$. I will now operate in a space of filters, as in the last section:

\begin{re-definition}\textsc{Space of Filters}
There exists a space of filters $\gamma$, comprising constituent filters $\Gamma \in \gamma$. $\Gamma_{i}$ represents the ith filter.
\end{re-definition}

I allow individuals to select which filter to apply to a menu, based on the competing factors I discuss in this section's preamble. Preference relations and related concepts will prove useful in setting up the environment.

\begin{definition}\textsc{Filter Preference Relation}
There exists a weak preference relation $\succsim_{\gamma}$ over the set of filters $\Gamma_{i} \in \gamma$.
\end{definition}

This preference relation $\succsim_{f}$ allows us to formally define the individual's preference over the filters $\Gamma_{i} \in \gamma$. The addition of choice over filters, as opposed to the exogenously-imposed consideration filter implicitly assumed earlier, necessitates the inclusion of this relation.

\begin{axiom}\textsc{Completeness of Filter Preferences}
The relation $\succsim_{\gamma}$ is complete. That is, for two filters $\Gamma_{i}$ and $\Gamma_{j}$ $\in \gamma$, $\Gamma_{i} \succsim_{\gamma} \Gamma_{j}$ or $\Gamma_{j} \succsim_{\gamma} \Gamma_{i}$.
\end{axiom}

This axiom states that the individual has a defined preference over every pair of filters with which they can be presented. A point to note is that I do not model uncertainty here: for simplicity, this model has full information over filters.

\begin{axiom}\textsc{Transitivity of Filter Preferences}
For any three filters $\Gamma_{i},\Gamma_{j}$, and $\Gamma_{k} \in \gamma$, $\Gamma_{i} \succsim_{\gamma} \Gamma_{j}$ and $\Gamma_{j} \succsim_{\gamma} \Gamma_{j}$ implies $\Gamma_{i} \succsim_{\gamma} \Gamma_{k}$.
\end{axiom}

Transitivity of $\succsim_{\gamma}$ prevents cycling, that is, preferences that move in a circular fashion. This is necessary as a consistency condition in order for choices to be indicative of underlying preferences, a key stipulation for utility representation.

\begin{axiom}\textsc{Rationality of Filter Preferences}
The relation $\succsim_{\gamma}$ is rational.
\end{axiom}

Rationality of $\succsim_{\gamma}$ follows from completeness and transitivity.\footnote{\cite{mas1995microeconomic} \textbf{Definition 1.B.1}.}

\begin{axiom}\textsc{Utility Representation}
For a menu $A$, there exists a utility representation $u_{f}$ over filters $\Gamma_{i} \in \gamma$ such that:

\begin{equation}
    \Gamma_{i} \succsim_{\gamma} \Gamma_{i} \Leftrightarrow u_{\gamma}(\Gamma_{i}) \geq u_{\gamma}(\Gamma_{i})
\end{equation}
\end{axiom}

By rationality, we know that there exists a utility representation.\footnote{\cite{mas1995microeconomic} \textbf{Definition 1.B.2}} This utility function is, by nature, ordinal in that that it captures preferences but not necessarily their intensity. This utility function, as well as the underlying preferences, hold the menu constant. That is, given a generic menu $A \in M(X)$, preferences are then well-defined over the space of filters $\gamma$. We can now work with the following generic function:

\begin{definition}\textsc{Filter Utility}
Individuals have filter preferences:
\begin{equation}
    u_\gamma = b_{\gamma}(\Gamma, A) - c_{\gamma}(\Gamma, A)
\end{equation}
\end{definition}

The utility function gives us the utility that the individual derives from applying a filter $\Gamma$ to menu $A$. The first argument, $b_{\gamma}(\Gamma, A)$, gives the benefit of the filter which I will define as coming from the alternative $x \in A$ that is eventually chosen. The cost, $c_{\gamma}(\Gamma, A)$, is the disutility of consideration.

\par This basic model fits in well the rational attention literature  in that it balances the benefit of choices with the costs of attention. Additionally sense, this model extension could be argued to be a motivation for why consideration filters are used in the first place: the standard model, in which individuals implicitly consider all goods, induces costs which may not be worthwhile.

\par One must note at this point that , to this point, this model has no empirical content. The terms $b_{\gamma}(\Gamma, A)$ and $c_{\gamma}(\Gamma, A)$ are too general to be identified. In order to impose some structure upon the model, I make additional specifications:

\begin{definition}\textsc{Choice of Alternative from Menu}
$c(\Gamma(A))$ selects the most preferred element from a menu $A$, according to the rational preference relation $\succsim_{x}$.
\end{definition}

$c(A)$ is simply the "best" alternative in $A$. This allows one to more specifically define the benefit of a particular filter, according to the alternative that it eventually generates. This makes sense as an individual will likely evaluate a consideration criterion according to the utility resulting from the choice that is eventually made.

\begin{definition}\textsc{Benefit of Consideration}
The benefit of consideration $b_{\gamma}(\Gamma, A)$ is, equivalently, $b_{\gamma}(c(\Gamma(A)))$.
\end{definition}

The benefit of a consideration filter $\Gamma$ is a function of the the best alternative in the generated consideration set, because that is the alternative which is chosen.

%according to some well-behaved function $f: \Gamma(A) \mapsto \mathbb{R}$, 

I now define the necessary elements in order to represent the cost of consideration.

\begin{definition}\textsc{Cardinality of Menu}
The cardinality of a menu $A$, $|A|$, the number of alternatives $x \in A$.
\end{definition}

Cardinality is the necessary concept to define the cost of consideration, as being a function of the cardinality of the consideration set.

\begin{definition}\textsc{Cost of Consideration}
The cost of consideration $c_{\gamma}(\Gamma, A)$ is, equivalently, $c_{\gamma}(|\Gamma(A)|)$.
\end{definition}

I define the disutility of consideration as direct function of the number of alternatives in the consideration set. This mirrors the well-known costly attention framework: there exists some cognitive cost of attention (time, mental strain, etc.) that induces negative utility coming from the sheer number of alternatives considered. I place more structure on the cost function:

\begin{axiom}\textsc{Convex Cost of Consideration}
The cost of consideration $c_{\gamma}(|\Gamma(A)|)$ is globally convex.
\end{axiom}

The more alternatives considered, the greater the disutility, and this disutility increases marginally: 
\begin{equation}
    \frac{\partial c(|\Gamma(A)|)}{\partial |\Gamma(A)|} > 0, \frac{\partial^2 c(|\Gamma(A)|)}{\partial |\Gamma(A)|^2} > 0
\end{equation}
\vspace{0.2cm}
Now that all arguments have been defined, I arrive at the following specification for the utility representation of preferences over filters:

\begin{definition}\textsc{Specified Filter Utility Function}
Individuals have filter preferences:
\begin{equation}
    u_{\gamma} = b_{\gamma}(c(\Gamma(A))) - c_{\gamma}(|\Gamma(A)|)
\end{equation}
\end{definition}

Filter utility $u_{\gamma}$ is broken down into two portions. As specified above, $uc(\Gamma(A)))$ denotes the utility derived from the alternative eventually chosen in accordance with rational preferences. A filter is only as good as the choice it leads to. The cost, $c_{\gamma}(|\Gamma(A)|)$ is a convex function of the number of alternatives an individual considers before making a decision.

\begin{definition}\textsc{Filter Choice Rule}
Individuals choose filters $\Gamma \in \gamma$ so as to maximize:
\begin{equation}
   c(\gamma) = \arg \max_{\Gamma \in \gamma} u_{\gamma} = b_{\gamma}(c(\Gamma(A))) - c_{\gamma}(|\Gamma(A)|)
\end{equation}
\end{definition}

This formally specifies the objective.

\begin{axiom}\textsc{Filter Choice Mandate}
$c(\gamma) \neq \emptyset$
\end{axiom}

The individual must choose a filter. This prevents the convex cost function from inducing null choice sets. Equipped with this model setup, I now present formal results.

\subsection{Results}

Below, I provide a remark on a cost-induced property of the filter choice process, as well two boundary results on the the number of alternatives an individual will consider.

\begin{definition}\textsc{Preference for Flexibility} A filter choice rule $c(\gamma)$ represents a preference for flexibility if $\Gamma_{1}(A)\supseteq\Gamma_{2}(A)$ implies that $c_{\gamma}(\Gamma_{1}, \Gamma_{2})=\Gamma_{1}$.
\end{definition}

Choice rules satisfying this Preference for Flexibility — a classic property in the decision theory literature — will induce individuals to choose filters that generate largest possible consideration sets. This is relevant in cases in which attention is costless, meaning individuals cannot be made worse off by more options given free disposal. Costly attention of the filter objective function leads Preference for Flexibility to fail in the model:

\begin{remark}
The filter choice rule $c_{\gamma}$ does not represent a preference for flexibility.
\end{remark}

The logic for this result follows directly from the existence of the cost function $c(|\Gamma(A)|)$. It may be the case that, while more alternatives may lead to better choices, the magnitude of the increased benefit may be outweighed by the cost of attention.

\vspace{0.2cm}

\par I now move into the two key theorems, detailing edge cases on filter choices.

\begin{theorem}\textsc{Costless Consideration Implies Full Consideration} \\
If $c_{\gamma}|\Gamma(A)|=0$ for all $\Gamma_i \in \gamma$, the individual considers all alternatives $x \in A$. $\Gamma(A) = A$.
\end{theorem}

\begin{proof}
See appendix.
\end{proof}

This result is intuitive. If attention is costless, then there is no reason to not consider all alternatives as the upside is potentially limitless with no downside cost.

\begin{theorem}\textsc{Worthless Consideration}
If $b_{\gamma}(\Gamma(A))$ is equal among all $\Gamma_i \in \gamma$, the individual chooses the filter $\Gamma_i \in \gamma$ so as to minimize $c_{\gamma}|\Gamma(A)|$.
\end{theorem}

\begin{proof}
See appendix.
\end{proof}

In Theorem 4, I shut down heterogeneity in cost by setting the cost function globally to zero. Theorem 5 can be seen as a reversal - here, eliminate heterogeneity in "rewards" by equating benefits from choices across all consideration sets. This result is similarly intuitive. If there is no potential benefit of a larger consideration set, the individual is justified in only considering one alternative, as there is assurance that they could not have improved their condition through increased consideration.\footnote{As a nod to contract theory, this result mirrors the well-known result that, in the classic principal-agent setting, setting equal wages for high output and low output will induce shirking on the part of the agent.} This result thus can be seen as a "no better off" theorem.

These two theorems complete my analysis of filter preferences by defining the edges of possible choices: at one extreme, individuals consider all alternatives available costlessly; at the other, individuals consider the minimum number of alternatives so as to simply satisfy the axiom that at least one alternative must be selected.

\subsection{Summary}

In this section, I extended the basic consideration model to a setting in which individuals may choose which consideration filters they apply to menus. In doing so, I provide a rational-attention model of limited consideration, modeling individuals as weighing the benefit of a larger consideration set with the convex costs of increased consideration. I provide two boundary results detailing cases in which an individual considers either all alternatives, or the minimum number possible. The next section outlines the viability of utility representation in a limited consideration setting.

\newpage

\section{Utility Representation}\label{sec:ur}

The ability to construct utility functions from observed choices — utility representation — relies on rational underlying preferences. This means that preferences must be complete and transitive. Under limited consideration, completeness naturally does not hold because the individual decision-maker does not necessarily consider every available alternative. Moreover, transitivity can also fail in the event that incomplete preferences lead choices to cycle.

\par As a result, limited consideration poses a fundamental threat to utility representation. In order to maintain the utility functions commonly assumed in structural models, assumptions need to be made on the process of consideration employed by individuals, who constitute the representative agents in applied literatures.

\par In this section, I begin by constructing a utility function that links to consideration-mediated choices. I then show that consideration filters that satisfy Independence of Others (IO) are sufficient for this form of utility representation. I then follow the approach of \cite{lleras2017more} by providing a modification of the weak axiom to match this consideration-consistent utility function.

\par In both the utility representation and weak axiom settings, it becomes clear that IO poses an extremely strong condition on choices and preferences. I acknowledge this and conjecture methods which can be used to weaken IO and preserve applicability.

\subsection{Consideration Utility Function}

I present a general utility function, which I will use for the results that follow. 

\begin{definition}\textsc{Generic Multi-Argument Utility}
\begin{equation}
   u(x \in A) = f(u_1(x), u_2(x),..., u_n(x))
\end{equation}
\end{definition}

This utility function maps each alternative $x \in A$ to the real numbers according to some amalgamation of $n$ different arguments. In the degenerate case there may exist only one argument. In order to capture the process of consideration, I denote the first constituent function, $u_1$, to be the \textit{threshold function}, the naming of which will become clear:

\begin{definition}\textsc{Threshold Function}
Within $f(u_1(x), u_2(x),..., u_n(x))$, $u_{1}$ is known as the threshold function.
\end{definition}

I require that, in order form some alternative $x \in A$ to be in the consideration set $\Gamma(A)$, the value generated by $u_1$ from $x$ must reach a certain threshold value, $k^*$:

\begin{axiom}\textsc{Threshold $k^*$}
    $x \in \Gamma(A)$ if and only if $u_1(x) \geq k^*$
\end{axiom}

Where $k$ finds its value among the real numbers:

\begin{axiom}\textsc{K is real} $k^* \in \mathbb{R}$
\end{axiom}

The consideration set from a menu $A$ thus consists of its alternatives that meet the threshold:

\begin{definition}\textsc{Threshold $k^*$ Consideration Set}
\begin{equation}
    \Gamma (A) = \{x \in A: u_{1} \geq k^{*}\}
\end{equation}
\end{definition}

The individual forms their consideration set from the alternatives $x \in A$ that meet a threshold condition that $u_{1}(x) \geq k^*$. This now provides a way to demonstrate consideration in the space of real numbers rather than purely through set theory. 

After completing this setup, the overall choice function now becomes:

\begin{definition}\textsc{Threshold Choice Function}
\begin{equation}
   c(A) = \arg \max_{x \in \Gamma (A)} f(u_1(x), u_2(x),..., u_n(x))
\end{equation}
\end{definition}

\begin{center}
    Where $\Gamma (A) = \{x \in A: u_{1} \geq k^{*}\}$
\end{center}

In words, the above choice function specifies the process:
\begin{enumerate}
    \item Individual is presented with menu $A$
    \item Individual narrows menu $A$ to consideration set $\Gamma(A)$ by only considering alternatives $x \in A$ for which $u_{1}(x) \geq k^*$
    \item An alternative $x$ is chosen from the consideration set $\Gamma(A)$ according to some rational preference relation.
\end{enumerate}

As stated above, this is one among many potential examples of how the set of real numbers can be used to facilitate the modeling of consideration. The setup I have develop allows me to present a utility representation result. I show that any choice process consistent with the above formulation must only involve consideration filters that satisfy IO:

\begin{theorem}\textsc{IO Utility Representation}
$\Gamma$ satisfies IO if and only if $\exists k^*$ and $u_1(x): X \mapsto \mathbb{R}$ such that $\Gamma (A) = \{x \in A: u_{1} \geq k^{*}\}$.
\end{theorem}

\begin{proof}
See appendix.
\end{proof}

Any filter used in the specified choice process must be IO. Recall that IO is equivalent to the joint presence of Sen's $\alpha$ and Condition $\tau$, and so these two conditions may also substitute into the result.

\par The idea of the proof, found in the appendix, is to fix the threshold $k^*$ to 1, and define $k^*$ as 1 for all alternatives within the IO-generated consideration set, and 0 for those without. By demonstrating that the set of alternatives meeting the threshold are also those within the consideration set, the proof is completed.

\par The above exercise shows an example of how consideration can be nested within utility function once properties of the consideration filters are specified. I now extend this exercise to the weak axiom, using IO as a proof of concept as to how the weak axiom can be modified to match the limited consideration setting.

\subsection{Weak Axiom Modifications}

The Weak Axiom of Revealed Preference (WARP) is a consistency condition that aligns choices with rational underlying preferences. The weak axiom, when it holds, mandates that if some alternative $x_1$ is chosen over $x_2$, then the reverse cannot happen when both are available in some other menu. In essence, preferences cannot "reverse" across two different menus. The weak axiom forms the basis for choice theory and, by extension, underpins utility representation.

\par Despite the ubiquity of the weak axiom, it does not naturally account for limited consideration. If, in the second menu, $x_1$ were not considered, then it is quite plausible for $x_2$ to be chosen, given that the individual may not even have been aware of the presence of $x_1$ in the menu. Therefore, the weak axiom needs to be modified to match the limited consideration setting.

\par This has been done before. \cite{lleras2017more} provide a modification of the weak axiom to account for the phenomenon of "choice overload." Choice overload refers to situations in which individuals consider certain alternatives in small menus, but somehow lose track of these alternatives when presented with a much larger menu that includes them. The rationale is that the overwhelming number of alternatives can be cognitively challenging and may induce forgetfulness or similar mental lapses. The choice overload WARP modification requires some simple notation. Refer to $S$ and $T$ as two menus within $M(X)$ and call $b$ an alternative in $X$. The choice overload WARP modification is:

\begin{axiom}\textsc{WARP Choice Overload (WARP-CO)}
For any nonempty S, there exists $b^*$ $\in S$ such that for any $T$ including $b^*$, if:

\begin{enumerate}
    \item $c(T) \in S$ and 
    \item $b^*=c(T^{'})$ for some $T^{'} \supset T$
\end{enumerate}
then $c(T)=b^*$
\end{axiom}

By requiring that the chosen alternative $b$ is considered in the larger set, WARP-CO "closes the hole" punctured by choice overload, allowing choices to again be consistent with rational preferences. Across many settings in which limited consideration may jeopardize completeness, it behoves the decision theorist to consider WARP modifications that are appropriate to the application of interest. As an example, I now provide a WARP modification that matches Independence of Others (IO), using the same notation as that of WARP-CO:

\begin{axiom}\textsc{WARP-IO}
For any nonempty S, there exists $b^*$ $\in S$ such that for any $T$ including $b^*$, if:

\begin{enumerate}
    \item C(S) = b*
    \item C(T) $\in$ S, and
    \item C(T) = b* if and only if c(Q) = b*, where Q = \{b*, x\}, $\forall$ x $\in$ T such that $\exists$ J $\in$ X such that x = C(J)
\end{enumerate}
then $c(T)=b^*$.
In addition:
\begin{enumerate}
    \item consider B $\subset$ X, where B = \{b, $\emptyset$\}
    \item if c(B) = $\emptyset$, then c(J) $\neq$ b for all menus J $\in$ X
\end{enumerate}
\end{axiom}

Recall the definition of IO:
\begin{re-definition}\textsc{Independence of Others}
A consideration filer $\Gamma$ satisfies Independence of Others (IO) if one of the following two conditions holds:
\begin{enumerate}
    \item \textit{x $\in$ $\Gamma (A)$ $\forall$ A $\in$ M(X) s.t. x $\in A$} \textbf{or}
    \item \textit{x $\notin$ $\Gamma (A)$ $\forall$ A $\in$ M(X)} 
\end{enumerate}
\end{re-definition}

The two conditions I provide in WARP-IO correspond to the two portions of the IO definition, respectively. In the first case, WARP-IO mandates that any choice must pairwise beat every other alternative in the menu. This matches full consideration in that there cannot be a case in which an alternative is selected despite not being preferred to some other alternative, which may happen if limited consideration restricts the scope of the consideration set. The second branch of WARP-IO concerns alternatives that are not chosen when they are the only alternative available. Naturally, this must mean that these alternatives, for some reason, were not considered, given that they are in the available set\footnote{I again remind the reader that, throughout the paper, I assume that the available set only includes alternatives that are within the individual's budget set.}. In this case, they are \textit{never chosen}, as IO states that alternatives that are not considered in some instance are never considered in any menu.

\par I have presented a modification to the Weak Axiom of Revealed Preference (WARP) to align with consideration heuristics that are modeled by IO filters. In doing so, I follow in the vein of \cite{lleras2017more}, who also devise a WARP modification to account for the nuances of consideration-mediated choices. Similar modifications are, in principle, possible for any property of consideration filters, and future literature can make large strides by developing the appropriate modifications to match the common behavioral processes most commonly observed in real-world economic settings.

\subsection{Summary}

In section, I have explored the implications of limited consideration on utility representation. Choices made under limited consideration may appear to reflect underlying preferences that are neither complete nor transitive, presenting a threat to the ability to use utility functions to represent consideration-mediated choices. In order to preserve utility representation, assumptions need to be made on the properties of consideration filters. As an example, I show an example of a utility function which captures choices made using an IO consideration filter, showing that IO is sufficient to model choices made via the objective function I set up.

\par I also address the Weak Axiom of Revealed Preference (WARP), which implies full consideration. I follow \cite{lleras2017more} by modifying the weak axiom to match a property of consideration filters, providing an example for IO. In future, as I discuss in the next section, the characterizations I provide ought to be extended to cover filter properties that are not as strong as IO.

\section{Future Directions}\label{sec:fd}

In this analysis I provide a language for discussing limited consideration, extended the basic model, and conjectured conditions for utility representation. Below I briefly mention three potential avenues for future work on limited consideration.
\vspace{0.2cm}

\textit{Weakening IO.} IO is clearly the strongest condition one can impose upon the formation of consideration sets — either an alternative is always considered when available, or else it is never considered, not even in the singleton set. IO, the foundation for some of the results in this paper, is clearly not flexible enough to match the nuances in individual behavior. However, the standard model, and classical revealed preference, make a similarly strong assumption: that the available set is always the consideration set ($\Gamma(A)=A$). Full consideration, a special case of IO,\footnote{Define the IO filter rule as: every alternative $x \in A$ is always considered when available. The available set is then always equivalent to the consideration set.} is therefore not realistic either, and so weakening of IO must be explored. This will require prudence, however, as the desired condition must be weaker than IO, while still having enough "bite" to ensure falsifiability.

\textit{Incorporating Consideration into Structural Models.} Many empirical phenomena could be better understood using the limited consideration framework. For example, \cite{larcom2017benefits} explore route choice in the London subway system before and after a temporary shortage, finding that some commuters used different routes after the transportation restart, meaning they were not optimizing before the strike. The authors build a structural model of route choice, finding that daily commuters often were not aware of routes that were faster than the ones they previously used. They allude to limited consideration, wondering why commuters did not experiment enough before the strike. One potential answer is limited consideration: individuals did not consider all available routes. This could take the form of a satisficing heuristic, or it could be modeled using the rational-attention model I developed in Section \ref{sec:prefs} in the event that the process of analyzing routes is seen to be costly. 

\textit{How Much Do We Toss Out?} Related to the first two suggestions, literature across all subfields assumes full consideration in some fashion or another. Given that this is not a realistic axiom, it is worth examining what sorts of theoretical and empirical results are no longer viable once one understands the salience of limited consideration. For example, in a setting in which individuals often use consideration heuristics, welfare analysis grounded in observed choices is likely to need adjustment.

\section{Conclusion}\label{sec:conc}

Revealed preference takes observed choices to be indicative of individual preferences. For example, if, given the set $\{A, B, ..., Z\}$, an individual chooses $R$, revealed preference indicates that $R$ is preferred to all other letters. This approach to choice theory assumes that individuals examine every available option before making a choice. In contrast, limited consideration posits that individuals narrow menus into consideration sets before making choices. This framework is better suited to modeling individual decision-making, which often involves various rules of thumb that filter out certain options.

\par The literature on limited consideration and related processes has been well-developed and includes theoretical models\footnote{\cite{masatlioglu_revealed_2012}; \cite{masatlioglu2015completing}; \cite{lleras2017more}}, consumer choice analyses\footnote{\cite{hauser1990evaluation}; \cite{roberts1991development}; \cite{erdem1996decision}}, axiomatic characterizations of normative preferences, \footnote{\cite{cherepanov2013rationalization} and \cite{ridout2021choosing}} and structural work in various empirical settings.\footnote{\cite{abaluck2016evolving}; \cite{larcom2017benefits}; \cite{abaluck2021consumers}}

\par In this paper, I provide a general model of limited consideration to unite the literature. I begin by outlining the main features of consideration model: individuals observe menus, narrow them into consideration sets, and make choices from said consideration sets. The channel by which menus are translated into consideration sets is captured by consideration filters, functions that downsize menus into consideration sets by mapping them to one of their subsets. Consideration filters correspond to various rules of thumb that individuals may use in narrowing down menus. To account for the large number of heuristics that individuals may use in practice, I introduce a number of properties that consideration filters may have. These properties describe the manner in which a menu begets a consideration set, and are neither mutually exclusive nor mandated to coexist with one another. A consideration filter may satisfy one, all, or none of the properties I introduce. The strongest condition, Independence of Others (IO), describes consideration filters which select a certain set of alternatives in any menu in which they appear, and never selects any alternative not in this set. IO, which closely approximates the standard rational model, forms the basis for a number of later results in the paper.

\par I then extend the consideration model in two ways. First, I develop a model of sequential consideration, which allows more than one filter to be applied to a given menu. This matches settings in which individuals are thought to apply more than one rule of thumb in narrowing down large choice sets. I introduce the concept of commutativity, borrowing from algebra to describe filters which can be applied to a menu in any order and still generate the same final consideration set. Filters satisfying IO are always commutative. My second model extension is a rational-attention analogue, in which I model an individual who must choose which filter to apply to a given menu. Individuals weigh competing forces in choosing a filter: the greater optionality associated with a larger consideration set and the examination costs associated with sifting through a large number of alternatives. I present two boundary results, showing in which cases an individual will consider every alternative, or the minimum number of alternatives.

\par I address the implications of limited consideration on utilty representation. The ability to construct utility functions corresponding to observed choices relies on underlying preferences being both complete and transitive. In the consideration model, both conditions often fail. To counteract this, I construct a utility function that accurately models choices made using an IO consideration filter. Such a link between consideration-based utility functions and the filter properties that may generate them may be possible for a large array of consideration heuristics. I also provide a modification to the Weak Axiom that corresponds to IO. In both cases, I use IO as a basic proof of concept to demonstrate how standard choice theory can be reconciled with the limited consideration framework.

\par There are many potential future directions for theoretical work on limited consideration. For example, IO can be weakened to find a filter property that is more realistic yet tractable. In addition, current structural choice models used by applied economists can be better reconciled with limited consideration, especially in empirical settings in which choices are thought to be made from consideration sets.

\par In summary, I have presented a detailed characterization of limited consideration, nesting some of the prior literature into a formal language while also developing novel model extensions. The hope is that a complete theory of consideration, building off this work as well as that of other scholars, will improve the robustness and applicability of rational choice theory.

\newpage

%Hello \cite{abaluck2016evolving} 
%Hello \cite{abaluck2016evolving}
%hello \cite{manzini2007sequentially}
%hello \cite{abaluck2021consumers}
%hello \cite{apesteguia2013choice}
%Hello \cite{caplin2011search}
\bibliographystyle{ecta}
\bibliography{main}

\newpage
\section*{Appendix: Proofs of Results in Main Text}\label{sec:app}

\subsection*{Theorem 1: Sen's $\alpha$ and Condition $\tau$ $\Leftrightarrow$ IO}

For the if direction, I show that any filter $\Gamma$ satisfying Sen's $\alpha$ and Condition $\tau$ is also IO. Suppose filter $\Gamma$ satisfies $\alpha$ and $\tau$. Further suppose that some alternative $x$ is in the consideration set generated by this $\Gamma$ on menu $A$. By Sen's $\alpha$, $x$ is in the consideration set of the singleton menu $\{x\} \in A$.\footnote{This is the going down step.} By $\tau$, $x$ is in the consideration set of any menu that includes $x$, since any such menu is a superset of $\{x\}$.\footnote{This is the going up step.} Since $x$ is always considered when available, $\Gamma$ is IO.

For the only if direction, I now show that any filter $\Gamma$ satisfying IO also satisfies Sen's $\alpha$ and Condition $\tau$. Suppose some filter $\Gamma$ satisfies IO. Further suppose that some alternative $x$ is in the consideration set generated by this $\Gamma$ on menu $A$. By IO, $x$ is always considered when available. By definition, $x$ appears in all subsets (Sen's $\alpha$)and supersets (Condition $\tau$) of $A$. Therefore $\Gamma$ satisfies Sen's $\alpha$ and Condition $\tau$, as desired.

\subsection*{Theorem 2: IO and Commutativity with 2 Filters}

I start with the if direction: if $\Gamma_{1}$ and $\Gamma_{2}$ are IO, then they are commutative for any menu $A \in M(X)$. This simply requires me to show that $x \in \Gamma_{12}(A)$ if and only if $x \in \Gamma_{21}(A)$. First, assume that $x \in \Gamma_{12}(A)$. Recall that, if an IO filter retains some alternative $x$, then it must retain $x$ in all menus $A$ that contain $x$. Therefore, $x \in \Gamma_{12}(A)$ implies that $x \in \Gamma_{1}(A)$ for if $x \in A$, and it is also true that $x \in \Gamma_{2}(A)$ for if $x \in A$. Now, recall that $\Gamma_{21}$ applies filter $\Gamma_1$ followed by $\Gamma_2$. Both filters, as I have shown, always retain $x$ when $x \in A$, and so $x \in \Gamma_{21}$.

Now, for the only if direction, which entails showing that if any two filters $\Gamma_{1}$ and $\Gamma_{2}$ are commutative for any menu $A$, then they must be IO. This result can be proved by inspection, noting the intuition behind IO. If in the event that filter $\Gamma_1$ is not IO, there necessarily exists a pathological case in which the consideration set generated by $\Gamma_1$'s involves a comparative selection process, i.e. alternatives are selected based on their desirability relative to others. In that case, the consideration set generated the successive application of $\Gamma_{1}$ and $\Gamma_{2}$ is clearly dependent upon the structure of the original menu.

\subsection*{Theorem 3: IO and Commutativity with \textit{N} Filters}

I prove this result using Theorem 2 as a base case. By Theorem 2, any two filters $\Gamma_1$ and $\Gamma_2$ are commutative for any menus $A \in M(X)$ if and only if they are IO. Because $\Gamma_1$ and $\Gamma_2$ are commutative, they can be collapse into one filter, since the order of their application does not matter. Call this new filter $\Gamma_{c}$. Suppose one adds a third filter $\Gamma_3$. By Theorem 2, $\Gamma_{c}$ and $\Gamma_3$ are commutative if and only if they are IO. Recalling that $\Gamma_c$ is an amalgam of $\Gamma_1$ and $\Gamma_2$, filters $\Gamma_1$, $\Gamma_2$, and $\Gamma_3$ are commutative if and only if they are IO.

If I add a fourth filter $\Gamma_4$, the same approach works by collapsing the first three filters into one. The proof strategy scales up for any $n$ filters.

%We can prove this result using our knowledge of \textbf{Theorem 4}. Since \textbf{Theorem 4} is true, we know that any two commutative filters are IO and vice versa. This is our "base case" for the $n$-step process. Afix an IO filter $\Gamma_{3}$ to the two menus, but consider the filter $\Gamma_{12} = \Gamma_{21}$ as one filter rather than two, which we can do due to commutativity. We can now apply \textbf{Theorem 4} since we have reduced the $n$-step problem to 2 filters. We can keep doing this for any $n$ number of steps.

\subsection*{Theorem 4: Costless Consideration Implies Full Consideration}

Recall the choice function

\begin{equation}
   c(\gamma) = \arg \max_{\Gamma \in \gamma} u_{\Gamma} = b_{\gamma}(c(\Gamma(A))) - c_{\gamma}(|\Gamma(A)|)
\end{equation}

If consideration is costless, $c_{\gamma}(|\Gamma(A)|) = 0$, therefore we have:

\begin{equation}
   c(\gamma) = \arg \max_{\Gamma \in \gamma} u_{\Gamma} = b_{\gamma}(c(\Gamma(A)))
\end{equation}

Which is maximized by considering all alternatives $x \in A$.\footnote{This assumes the benefit of consideration $b_{\gamma}$ is non-decreasing.}

%Given any amount of considered alternatives $|\Gamma(A)|$, there is no cost to consider more. Call the "marginal considered alternative" $m$, that is, the alternative not currently considered but that could be. So long as there is some positive probability that $m>sup(\Gamma(A))$, the individual ought to add it to the consideration set. The individual will then consider all alternatives.

\subsection*{Theorem 5: Worthless Consideration}

Recall the choice function

\begin{equation}
   c(\gamma) = \arg \max_{\Gamma \in \gamma} u_{\Gamma} = b_{\gamma}(c(\Gamma(A))) - c(|\Gamma(A)|)
\end{equation}

If the benefit of consideration $b_{\gamma}(c(\Gamma(A)))$ is constant across all filters, then it is invariant to any change in the filter and this has no effect on the optimal choice. The problem reduces to cost minimization:

\begin{equation}
   c(\gamma) = \arg \max_{\Gamma \in \gamma} u_{\Gamma} = -c(|\Gamma(A)|)
\end{equation}

%However, with constant $sup(\Gamma(A))$ across all filters $\Gamma \in \gamma$, the benefit is constant. 

In order to maximize the objective, while satisfying \textbf{Axiom 6}, the filter choice mandate, the individual will simply choose the filter that minimizes the cost of consideration.

\subsection*{Theorem 6: IO Utility Representation}

%The if direction, that IO filters can be represented by the threshold choice function, is straightforward. Father all alternatives in 

%The proof strategy  

%, D

The if direction, that IO filters can be represented by the threshold choice function, is straightforward. List all alternatives $x \in A$ which are in the consideration set of the IO filter $\Gamma$. For each $x \in \Gamma(A)$, set $u_1(x) = 1$. For each $x \notin \Gamma(A)$, set $u_1(x) = 0$. Then set $k^* = 1$. Therefore, all alternatives $\Gamma_1$ meet the $u_1$ threshold.

The only if direction makes us of the same technique, paired with the application Sen's $\alpha$ and Condition $\tau$ to the relevant menus. To prove this direction. Define a set $Y\subseteq X$ as follows
	$$y\in Y \text{ iff } x\in \Gamma(A) \text{ for some } A.$$
	Define 
	\begin{equation}
		u_1(x)=\begin{cases}
			1 & \text{ if } x\in Y\\
			0 & \text{ if } x\notin Y
		\end{cases}
	\end{equation}
	
	It remains to verify that $\Gamma(A)=\{x\in A: u_1(x)\geq k^*\}$ as desired. Let's check. By construction, the right hand side equals $\{x\in A: x\in Y\}=A \cap Y$. So it remains to check whether $\Gamma(A)=A\cap Y$. There are two arguments needed:
	\begin{enumerate}
		\item If $x\in \Gamma(A)$ then $x\in A \cap Y$
		\item If  $x\in A \cap Y$ then $x\in \Gamma(A)$. 
	\end{enumerate}
	
	To prove the first argument, assume that $x\in \Gamma(A)$. Then automatically $x\in A$, and  $x\in Y$ By definition of $Y$.
	
	To prove the second argument, assume that $x\in A \cap Y$. Then by definition of $Y$ there exists a set $B$ such that $x\in \Gamma(B)$. Let's consider the set $C:=A\cap B$. We know $x\in C$ since both $A$ and $B$ contain it. First, apply Sen's $\alpha$ condition to $x\in C\subseteq B$. This implies that $x\in \Gamma(C)$. Then apply Condition $\tau$ condition to $x\in C \subseteq A$. This implies that $x\in \Gamma(A)$, as desired.

%\bibliographystyle{my-style}
%\bibliographystyle{ecta}
%\bibliography{references}

% \addtocontents{toc}{\setcounter{tocdepth}{-1}}%removes appendices from table of contents

\end{document}